\documentclass[11pt,a4paper]{article}

\usepackage{jcappub}
\usepackage[margin=0.95in]{geometry}

\usepackage[utf8]{inputenc}
\usepackage{graphicx}
\usepackage{tensor}
\usepackage{verbatim}
\usepackage{epsfig}
\usepackage{subfigure}
\usepackage{color}
\usepackage{multirow}
\usepackage{comment}
\usepackage{slashed}
\usepackage{amsmath}
\usepackage{ulem}
\usepackage{framed}
\usepackage{amssymb,amsmath,amsfonts,mathrsfs}
\usepackage[dvipsnames]{xcolor}
\usepackage{longtable}
\usepackage[mathscr]{euscript}
\usepackage{framed}
\usepackage{mdframed} 

\usepackage{cancel}
\usepackage{hyperref}
\hypersetup{colorlinks, citecolor=bluscuro, linkcolor=black, urlcolor=bluscuro}
\definecolor{rossos}{cmyk}{0,1,1,0.55}
\definecolor{bluscuro}{rgb}{0.15, 0.2, .85}
\definecolor{bluchiaro}{cmyk}{1,.3,0.,0.1}

\numberwithin{equation}{section}
\renewcommand\theequation{\arabic{section}.\arabic{equation}}

\usepackage{tikz}

\usepackage{tcolorbox}

\newcommand{\llp}{\left [}
\newcommand{\rrp}{\right ]}

\newcommand{\lp}{\left (}
\newcommand{\rp}{\right )}

\newcommand{\vx}{\vec{x}}

\def\PBH{\text{\tiny  PBH}}

\newcommand{\be}{\begin{equation}\begin{aligned}}
\newcommand{\ee}{\end{aligned}\end{equation}}
\newcommand{\bbe}{\begin{align}}
\newcommand{\eee}{\end{align}}
\newcommand{\bea}{\begin{eqnarray}}
\newcommand{\eea}{\end{eqnarray}}
\def\beq{\begin{equation}}
\def\eeq{\end{equation}}

\def\lsim{\mathrel{\rlap{\lower4pt\hbox{\hskip0.5pt$\sim$}}
		\raise1pt\hbox{$<$}}}     
\def\gsim{\mathrel{\rlap{\lower4pt\hbox{\hskip0.5pt$\sim$}}
		\raise1pt\hbox{$>$}}}     
	
	\def\d{{\rm d}}	
\newcommand{\arXiv}[2]{\href{http://arxiv.org/pdf/#1}{{\tt [#2/#1]}}}
\newcommand{\arXivold}[1]{\href{http://arxiv.org/pdf/#1}{{\tt [#1]}}}

\def\npbh{\bar n_\text{\tiny PBH}}

\def\mpbh{M_\text{\tiny PBH}}

\def\form{\text{\tiny form}}

\def\fpbh{f_\text{\tiny PBH}}
\def\rhopbh{ \rho_\text{\tiny PBH}}

\title{The  Clustering Evolution \\ of Primordial Black Holes}
\author[a]{V. De Luca,}
\author[b]{V. Desjacques,}
\author[a]{G. Franciolini,}
\author[a,c]{A. Riotto}

\affiliation[a]{
	Department of Theoretical Physics and Center for Astroparticle Physics (CAP) \\
			24 quai E. Ansermet, CH-1211 Geneva 4, Switzerland}
\affiliation[b]{Physics department and Asher Space Science Institute, Technion, Haifa 3200003, Israel}
\affiliation[c]{INFN, Sezione di Roma, Piazzale Aldo Moro 2, 00185, Roma, Italy}

\abstract{
	Primordial black holes  might comprise a significant fraction of the dark matter in the Universe and be responsible for the 
	gravitational wave signals from black hole mergers observed by the LIGO/Virgo collaboration. 
	The spatial clustering of primordial black holes might affect their merger rates and have a significant impact on the constraints on their masses and abundances.
	We provide some analytical treatment of the primordial black hole spatial  clustering evolution,  compare our results with some of the existing N-body numerical simulations  and discuss the implications for the black hole merger rates. If primordial black holes  contribute to a small  fraction of  the dark matter, primordial black hole clustering is not relevant.  
On the other hand, for  a large  contribution to   the dark matter, we argue that the  clustering may  increase the late time Universe merger rate to a level compatible with the LIGO/Virgo detection rate.  As for the early Universe merger rate of black hole  binaries formed at primordial epochs,  clustering alleviates the LIGO/Virgo constraints, but does not evade them.
}
\emailAdd{valerio.deluca@unige.ch}
\emailAdd{dvince@physics.technion.ac.il}
\emailAdd{gabriele.franciolini@unige.ch}
\emailAdd{antonio.riotto@unige.ch}
\begin{document}

\maketitle
\flushbottom

\section{Introduction}
\noindent

The  LIGO/Virgo collaboration has observed several gravitational wave (GW) signals coming from the coalescence of massive black holes (BHs) during the first three observational runs  \cite{LIGOScientific:2018mvr,LIGOScientific:2020stg, Abbott:2020khf}. These observations have renewed the interest in the hypothesis that BHs are of primordial origin and formed in the early Universe, see Ref.~\cite{revPBH}  for a review. In particular, the emphasis is on the possibility that the Primordial Black Holes (PBHs) comprise a significant fraction $f_\PBH$ of the dark matter (DM) in the Universe \cite{Sasaki:2016jop,Bird:2016dcv}.

The evolution and the survival of PBH binaries during the history of the Universe until their merger is a key ingredient of the calculation of the gravitational wave signal. In particular, PBH binaries might be disrupted by close encounters with a third PBH. While isolated PBH binaries are likely not affected by three-body encounters, PBH binaries residing in PBH clusters might undergo such interactions more frequently and, thus, the expected merger rate might be modified.
Therefore, it is clear that the extent to which PBHs cluster is an important question to address, especially since conflicting results about the impact of clustering onto PBH merger rates have been presented in the literature \cite{Vaskonen:2019jpv,j1}. A significant PBH clustering might also help in modifying constraints arising from microlensing observations and from the cosmic microwave background, see for instance \cite{Garcia-Bellido:2017xvr} and Ref.~\cite{revGreen} for a recent review.

Even though PBHs, which form at horizon re-entry through the collapse of large overdensities produced during inflation \cite{revPBH}, are initially not clustered \cite{cl1,cl2,cl3,cl4}, PBH clusters form not long after matter-radiation equality \cite{raidal,inman} if $f_\PBH$ is large.
While N-body simulations will likely have a final say with regards to PBH clustering, analytical insight is certainly useful for understanding the complex evolution of the PBH population.

The purpose of this paper is to offer preliminary considerations on the PBH clustering, and compare analytical approximations to the only fully cosmological  existing N-body simulation \cite{inman} and to some numerical results in Ref.  \cite{raidal}. \footnote{For a numerical study of the dynamics of a single PBH cluster, see also Refs. \cite{j1,Trashorras:2020mwn}. }

Our findings indicate that, on the one hand, the merger rate of PBH binaries formed in the early Universe decreases in the presence of clustering, but it still remains above the current LIGO/Virgo observed detection rate. On the other hand, we find that the merger rate of PBH binaries formed in the late Universe is increased by PBH clustering when $f_\PBH$ is of order unity. This result is important because, in the absence of clustering, the estimated late-time merger rate is orders of magnitude below the current detection rate and, therefore, usually neglected in phenomenological studies.

Our paper is organised as follows. After some preliminary definitions in Section 2, we discuss the initial conditions for the PBH power spectrum in Section 3 and the linear, quasi-linear and non-linear regimes in Sections 4, 5 and 6, respectively. The role of evaporation is discussed in Section 7, the impact of clustering on the merger rates in Section 8 and we conclude in Section 9. Appendix A   contains some considerations about the clustering of an extra dark matter component in the presence of PBHs, while Appendix B offers comments about the description of a BH in an expanding universe.

\section{Some definitions}
\noindent
PBHs form if energy density perturbations are sizeable enough when the corresponding wavelengths are re-entering the horizon (after inflation).
Such large density fluctuations collapse to form PBHs almost immediately after horizon re-entry \cite{revPBH}. The resulting PBH mass is of the order of the mass contained in the corresponding horizon volume.

We are interested in the properties of the PBH spatial distribution at any subsequent redshift $z$.
To characterize the PBH two-point correlation function (or, simply, correlation function) as a function of the comoving separation $x=|\vx|$, we introduce the overdensity of discrete PBH centers at position $\vx_i$ with respect to the total background DM energy density,
\begin{eqnarray}
\frac{\delta\rho_\PBH({\vec x},z)}{f_\PBH\overline\rho_{\text{\tiny DM}}}=\frac{1}{\npbh}\sum_i \delta_{\text{\tiny D}}(\vx-\vx_i(z))-1,
\end{eqnarray}
where $\delta_{\text{\tiny D}}(\vx)$ is the three-dimensional Dirac distribution, and 
\be
\npbh\simeq 3.2 \,\fpbh\,\left(\frac{20\,M_\odot/h}{ \mpbh}\right)(h/{\rm kpc})^{3}
\ee
is the average number density of PBH per comoving volume. Here, $i$ runs over the  positions of PBHs.
The two-point correlation function of this discrete point process takes the general form
\begin{eqnarray}
\Big< \frac{\delta\rho_\PBH({\vec x},z)}{\overline\rho_{\text{\tiny DM}}}\frac{\delta\rho_\PBH(0,z)}{\overline\rho_{\text{\tiny DM}}} \Big>=
\frac{f^2_\PBH}{\npbh}\delta_{\text{\tiny D}}(\vx)+ \xi(x,z).
\label{eq:PBH2pt}
\end{eqnarray}
This expression emphasizes that $\xi(x,z)$ is the so-called reduced PBH correlation function and, thus, is distinct from the additive Poisson noise proportional to the Dirac delta. Characterizing  the magnitude and evolution of the reduced correlation function is the goal of this paper. The corresponding PBH  power spectrum  
\begin{eqnarray}
\Delta^2(k,z)= \frac{k^3}{2 \pi^2}\int \d ^3 x\,  e^{i {\vec k} \cdot {\vec x}}\, \Big< \frac{\delta\rho_\PBH({\vec x},z)}{\overline\rho_{\text{\tiny DM}}}\frac{\delta\rho_\PBH(0,z)}{\overline\rho_{\text{\tiny DM}}} \Big>,
\end{eqnarray}
is conveniently defined relative to the total cold dark matter average density.

\section{Initial conditions }
\noindent
Generally speaking, hierarchical clustering implies that, below a characteristic PBH clustering length, fluctuations in PBH number counts are dominated by the reduced PBH correlation function $\xi (x,z)$, while Poisson fluctuations dominate on larger scales. If PBHs initially form from the collapse of sizeable fluctuations upon horizon re-entry, the
exact value of the initial clustering length is, in principle, sensitive to the shape of the primordial curvature power spectrum. However at formation, and rather irrespectively of the shape of the power spectrum and the PBH mass function, the characteristic PBH clustering length is significantly smaller than the mean comoving PBH separation, rendering clustering not relevant at the time of formation \cite{cl1,cl2,cl3,cl4}. This is true in the absence of primordial non-Gaussianity correlating long- and short-wavelength fluctuations, which we assume from now on. Therefore, we can assume a Poisson distribution at the formation redshift $z_{\text{\tiny i}}$ and approximate the initial PBH power spectrum as \cite{inman}\footnote{As for the  way PBHs are distributed in mass, we consider here a single PBH mass. This is a good approximation not only if the power spectrum of the curvature perturbation is peaked around a single comoving momentum, but also when it is broad. Indeed, in such a case, the  mass function is peaked at the smallest PBH which can be formed upon horizon re-entry \cite{broad}.}
\begin{eqnarray}
\Delta^2_{\text{\tiny i}}(k)&=&\frac{k^3}{2 \pi^2}\int \d ^3 x\,  e^{i {\vec k} \cdot {\vec x}} \Big< \frac{\delta\rho_\PBH({\vec x},z_{\text{\tiny i}})}{\overline\rho_{\text{\tiny DM}}}\frac{\delta\rho_\PBH(0,z_{\text{\tiny i}})}{\overline\rho_{\text{\tiny DM}}} \Big>
\approx \fpbh^2\left(\frac{k}{k_*}\right)^3,
\end{eqnarray}
where the characteristic wavenumber
\be
k_*=(2\pi^2 \npbh)^{1/3}\simeq 4\,\fpbh^{1/3}\, \left(\frac{20\,M_\odot/h}{\mpbh}\right)^{1/3} h/{\rm kpc}
\ee
is inversely proportional to the mean separation between PBHs.

\section{The linear regime }
\begin{figure}
	\centering
	\includegraphics[width= .7\linewidth]{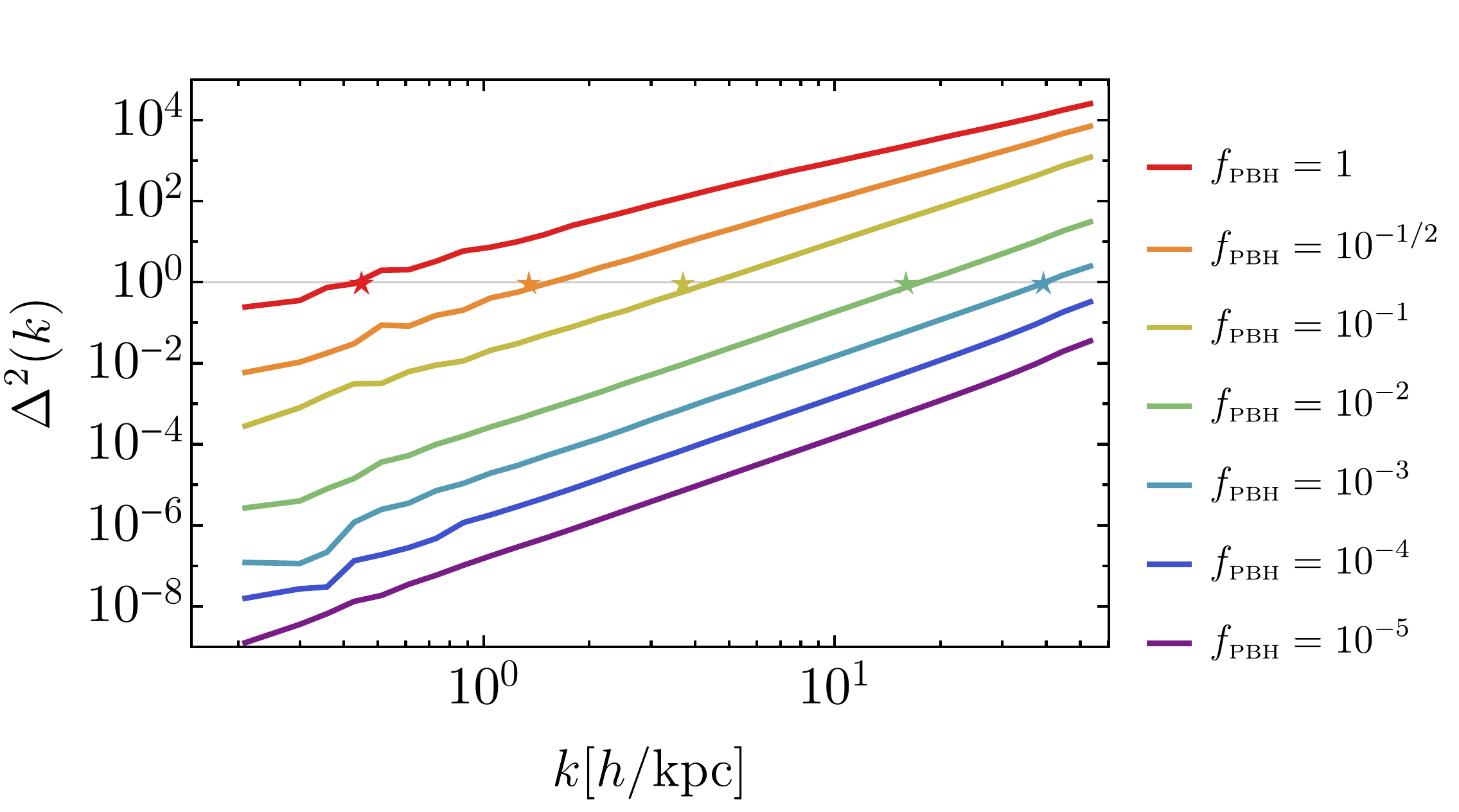}
	\caption{The PBH power spectra at $z=99$ for different values of $\fpbh$. The data are taken from from Ref.~\cite{inman}. The stars mark the transition to the quasi-linear regime as predicted by Eq.~(\ref{LL-QL}).}
	\label{fig1}
\end{figure}

\noindent
In the linear regime, the PBH density contrast is essentially frozen until matter-radiation equivalence, and subsequently grows linearly according to \cite{inman}
\be
\label{linear}
\Delta_{\text{\tiny L}}^{2}(k,z)\simeq\left(1+ \frac{3}{2}\fpbh\frac{1+z_\text{{\tiny eq}}}{1+z}\right)^2 \Delta^2_{\text{\tiny i}}(k),
\ee
in which we adopt the matter-dominated epoch behaviour  $(1+z)^{-1}$ for simplicity.
Linear (L) perturbations in the PBH number density enter the quasi-linear (QL) regime when the density contrast is of order unity.
This transition occurs at a different redshift depending on the wavenumber $k$ of the fluctuation. 
This happens approximately when 
\be
\Delta_{\text{\tiny L}}^{2}(k=k_{\text{\tiny{L-QL}}}(z),z)\simeq 1,
\ee
which implies
\be
k_{\text{\tiny L-QL}}(z)
\simeq \frac{4}{\fpbh^{1/3}}\left(\frac{20\,M_\odot/h}{\mpbh}\right)^{1/3} \left[1+ 26\fpbh \lp \frac{100}{1+z} \rp\right]^{-2/3} h/{\rm kpc}.
\label{LL-QL}
\ee
For illustration, we display in Fig.~\ref{fig1} the PBH power spectra extracted from the N-body simulations of Ref.~\cite{inman} at $z=99$  and kindly provided to us by the authors. The stars indicate the corresponding values of $k_{\text{\tiny{L-QL}}}$, which fit rather well the numerical results.

\section{The quasi-linear regime }
\noindent
Once the PBH perturbations leave the linear regime, they decouple from the Hubble flow, collapse and virialize to form halos with a virial density about 200 times the background density at the time of virialization. The formation of PBH halos is hierarchical like in a standard CDM cosmology: the small mass PBH halos which form first are the progenitors of more massive halos virializing at a later epoch. Before entering the fully non-linear regime, PBH perturbations experience a quasi-linear regime during which the power spectrum $\Delta^2$ at a given wavenumber $k$ grows from unity until $\sim 200$ (while the linearly evolved PBH averaged two-point correlator defined below grows from unity until 5.85 \cite{pad}). The collapse epoch is determined from the requirement that the (integrated) overdensity within a given comoving radius $R$ reaches a critical value $\delta_c\simeq 1.68$. Therefore, one shall consider the volume averaged correlation function 
\begin{eqnarray}
\overline{\xi}(R,z)&=&\frac{3}{4\pi R^3}\int_0^R\,\d s\, 4\pi s^2 \xi(s,z),
\nonumber\\
\xi(x,z)&\simeq & \int \frac{\d k}{k}\,  e^{i {\vec k} \cdot {\vec x}} \Delta^2(k,z),
\end{eqnarray}
which may be interpreted as a characteristic squared overdensity within the radius $R$.
In the second equation, the approximate sign emphasises that the large-scale Poisson piece in the correlation function is subdominant in the quasi-linear regime.
The volume averaged correlation function can also be thought of as measuring the power at some effective wavenumber
\be
\Delta^2(k,z)\simeq \overline{\xi}(1/k,z).
\ee
In order to give a prescription which connects the quasi-linear correlation function to the linear theory, one can use the conservation of particle pairs to write down an 
equation satisfied by $\xi(x,z)$. From the mean number of neighbours \cite{book}
\be
N(x,z)=\npbh 
\int_0^x\d s\, 4\pi s^2\left[1+\xi(s,z)\right],
\ee
and momentarily neglecting two-body relaxation along with evaporation from the PBH cluster (we will come back to these issues later on), the conservation of neighbours implies the equation \cite{book}
\be
\label{xievo}
\frac{\partial \xi}{\partial t}+ \frac{1}{a x^2}\frac{\partial}{\partial x}\left[x^2(1+\xi) v \right] = 0,
\ee
where $a$ is the scale factor and $v (x,t)$ denotes the mean relative velocity of pairs at separation $x$ and time $t$. This pair conservation equation yields the (mass conservation) relation \cite{book}
\be
x^3 (1 + \overline{\xi}) = R^3,
\ee
in terms of the initial shell radius $R$, from which one deduces that, as long as the evolution is linear, $ \overline{\xi} \ll 1$ and $R \sim x$ whereas, as clustering develops, $ \overline{\xi}$ increases and the scale $x$ becomes smaller than $R$. This implies that the correlation function in the quasi-linear regime $\overline{\xi}_\text{\tiny QL} (x)$ can be expressed in terms of the linear regime expression given by $\overline{\xi}_\text{\tiny L}(R)$. To spell out this relation, one considers a region surrounding a density peak in the linear stage, around which one expects clustering to take place \cite{pad}. The density profile around a peak is proportional to the underlying correlation function \cite{bbks} (ignoring the gradient contribution). Therefore, the linear integrated squared density contrast scales with the initial shell radius $R$ as $\overline{\xi}_\text{\tiny{L}}(R)$ so long as linear theory is valid. According to the standard spherical collapse model, such a perturbation expands to a maximum radius $x_{\text{\tiny{max}}}$ proportional  to $R/\overline{\xi}_\text{\tiny{L}}(R)$ \cite{book}. Taking the effective radius proportional to  $x_\text{\tiny{max}}$ and considering a halo of mass $M$, we have
\begin{eqnarray}
\overline{\xi}_{\text{\tiny QL}}(x)&\sim& 
\frac{M}{x^3}\sim \frac{R^3}{(R/\overline{\xi}_\text{\tiny{L}}(R))^3}
\sim \overline{\xi}_{\text{\tiny{L}}}^3(R),
\nonumber\\
R^3&\sim &x^3 \overline{\xi}_{\text{\tiny L}}^3\sim x^3\overline{\xi}_{\text{\tiny QL}}.
\end{eqnarray}
Since for the PBH we do have $\overline{\xi}_\text{\tiny{L}}(x)\sim x^{-3}$, we immediately obtain the scaling $\overline{\xi}_\text{\tiny{QL}}(x)\sim  x^{-9/4}$ or
\begin{eqnarray}
  \Delta_\text{\tiny{QL}}^2 (k) &\simeq& \left(\frac{k}{k_\text{\tiny{L-QL}}(z)}\right)^{9/4} \nonumber \\
  &\simeq& 0.04\,\fpbh^{3/4}
\left(\frac{20\,M_\odot/h}{\mpbh}\right)^{-3/4} \left[1+ 26\fpbh \lp \frac{100}{1+z} \rp\right]^{3/2} \left(\frac{k}{h/{\rm kpc}}\right)^{9/4}.
\label{L-QL}
\end{eqnarray}
This prediction is in good agreement with the data as shown in Fig.~\ref{fig2}.

The time dependence can be found upon assuming that the total DM provides the total energy density of the Universe (which is a good approximation until dark energy dominates). In such a case, the evolution has to be self-similar if the initial power spectrum is a power law \cite{book} and
the Boltzmann equation for the self-gravitating PBHs admits a self-similar solution of the form $\xi(x,t)=f(x/t^\alpha)$ (for more details, see Ref.~\cite{sb}). This solution is consistent with the linear behaviour of the correlation function for $\alpha=4/9$ only. Since the quasi-linear correlation function
$\overline{\xi}_\text{\tiny{QL}}(x,z)$ can only depend upon the combination $x(1+z)^{2/3}$, $\Delta^2_\text{\tiny QL}(k)$ must scale like
$(1+z)^{-3/2}$, a time dependence weaker than in the linear regime. This scaling is also apparent in Eq.~(\ref{L-QL}) if one considers the regime $26\fpbh(10^2/(1+z))\gsim 1$.

\section{The non-linear regime }
\noindent
To track the PBH perturbations in the non-linear regime, we rely on the stable clustering hypothesis which states that, although the separation between clusters is altered by the expansion of the Universe, their internal structure remains constant with time (i.e. they do not expand).

Under the stable clustering hypothesis, the pair conservation equation in the non-linear regime $\xi\gg 1$ can be recast into \cite{book}
\be
\label{xievo2}
\frac{\partial}{\partial t}(1+\xi)=\frac{H}{x^2}\frac{\partial}{\partial t}\left[a^3(1+\xi)\right]
\ee
and admits a power-law solution of the form 
\be
\label{xinlplaw}
\xi_{\text{\tiny{NL}}}(x,z)\sim \frac{x^{-m}}{(1+z)^{3-m}}.
\ee
As above, the index $m$ can be determined from self-similarity considerations, which imply that the correlation function must be of the form $\xi(x,t)=f(x/t^\alpha)$ with $\alpha=4/9$ to consistently reproduce the linear behaviour or, equivalently, $\xi(x,z)\sim f(x (1+z)^{2/3})$. Therefore, Eq.~\eqref{xinlplaw} shows that we must take $\xi_\text{\tiny NL}(x,z)\sim \big(x (1+z)^{2/3}\big)^{-m}$ and, consequently, $m=9/5$.
Again, we have assumed that the total DM density contributes to all the energy budget of the Universe. 

\begin{figure}[t!]
	\centering
	\includegraphics[width= .7\linewidth]{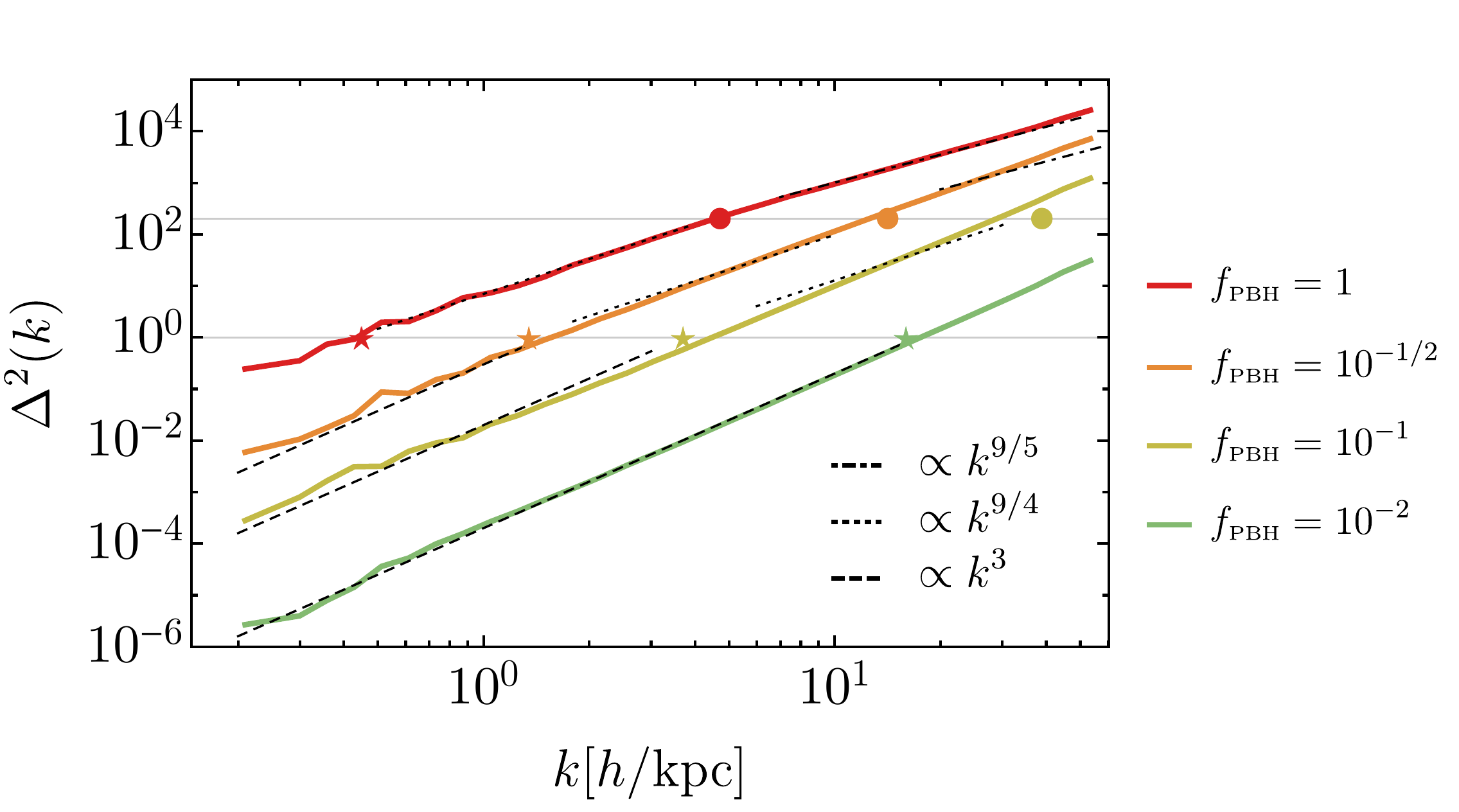}
	\caption{The PBH power spectra at $z=99$ for different values of $\fpbh$ from~\cite{inman}. The dots indicate the predictions of the expression (\ref{nl}) and the power laws are indicated by straight lines.}
	\label{fig2}
\end{figure}

The transition between the  quasi-linear and non-linear regime can be found upon requiring $\bar \xi \sim 200 $, or 
$(k_{\text{\tiny{L-QL}}}/k_{\text{\tiny{QL-NL}}})^{-9/4} \sim  200$. This gives
\begin{eqnarray}
k_{\text{\tiny{QL-NL}}}(z)  
\simeq 42  \fpbh ^{-1/3} \left( \frac{\mpbh}{20 M_\odot/h}\right)^{-1/3} \left[1+ 26 \fpbh \left( \frac{100}{1+z} \right) \right]^{-2/3} h/{\rm kpc}.
\label{nl}
\end{eqnarray}
The corresponding power spectrum thus reads 
\be
\Delta_{\text{\tiny{NL}}}^2 (k)\simeq 200 \left(\frac{k}{k_{\text{\tiny{QL-NL}}}(z)}\right)^{9/5}\simeq 0.2\,\fpbh^{3/5}\left( \frac{\mpbh}{20 M_\odot/h}\right)^{3/5}\left[1+ 26 \fpbh \left( \frac{100}{1+z} \right) \right]^{6/5} \left(\frac{k}{h/{\rm kpc}}\right)^{9/5},
\ee
which scales with redshift as $(1+z)^{-6/5}$ for $26\fpbh(10^2/(1+z))\gsim 1$. This prediction is compared in Fig.~\ref{fig2} to the numerical data at $z=99$ provided by the authors of Ref.~\cite{inman}. The dots indicate the position of $k_{\text{\tiny{QL-NL}}}$, while the straight lines represent the various power-laws expected in the quasi-linear and in the non-linear regime. Our findings are in fairly reasonable agreement with the numerical N-body data, especially given that we do not take into account the backreaction of the other DM component when $\fpbh\ll 1$. For $\fpbh=1$, this feedback is absent and the agreement between our prediction and the data is very good.

\begin{figure}[t!]
	\centering
	\includegraphics[width= .7\linewidth]{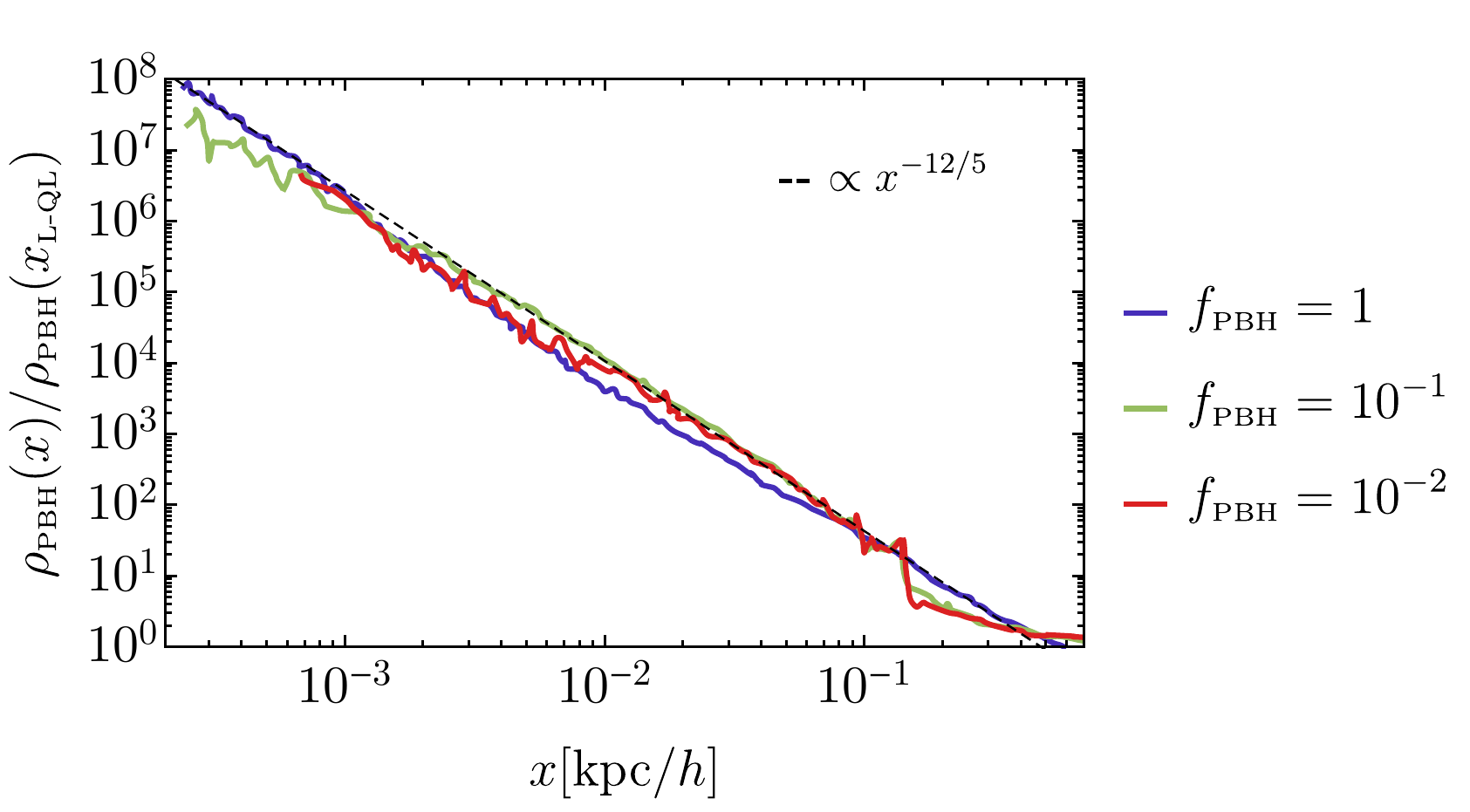}
	\caption{The PBH profile  at $z\simeq 1100$ for different values of $\fpbh$ obtained from Fig.~7 of Ref.~\cite{raidal}, together with our prediction (\ref{profile}) in dashed lines.}
	\label{fig3}
\end{figure}

We can further check the validity of our prediction as follows.
If we restrict ourselves to small scales, both members of a PBH pair are almost certainly drawn from the same PBH halo. In this limit, if the PBH density  profile is $\rhopbh(x)\sim x^{-\epsilon_\text{\tiny PBH}}$, then the two-point correlation function must behaves like $\sim x^{-2\epsilon_\text{\tiny PBH}+3}$ \cite{peebles,MS} as it is proportional to the square density profile. Imposing $(-2\epsilon_\text{\tiny PBH}+3)=-9/5$, we infer that the PBH density profile should satisfy
\be
\rhopbh(x)\sim x^{-12/5}.
\label{profile}
\ee
In Fig.~\ref{fig3}, we plot the numerical data shown in Fig.~7 of Ref.~\cite{raidal}, which illustrates the properties of the PBHs surrounding a {\it central} binary at $z\simeq 1100$ and must be interpreted as the PBH density profile rather than a correlation function as stated in Ref.~\cite{raidal}\footnote{We thank M. Raidal and H.~Veerm\"{a}e for clarifying discussions about  Fig.~7 of Ref.~\cite{raidal}.}. The density profile is the (conditional) correlation function subject to the constraint that one member of each particle pair certainly is at the center of a halo, while the (unconditional) correlation function is obtained when the location of each pair member is unconstrained. Our results fit well the PBH density profile found in Ref.~\cite{raidal}.

Another interesting question we can ask is what are the typical halos giving the largest contribution to the PBH correlation function. Applying Press-Schechter theory \cite{Press:1973iz} to an initial  Poisson power spectrum, the resulting number density of PBH halos with mass between $M$ and $(M+\d M)$ reads
\be
\label{halo function}
\frac{\d n(M,z)}{d M} =\frac{\overline{\rho}_{\text{\tiny{PBH}}}}{\sqrt{\pi}}\left(\frac{M}{M_*(z)}\right)^{1/2}\frac{e^{-M/M_*(z)}}{M^2},
\ee
where $\overline{\rho}_{\text{\tiny{PBH}}}$ is the average PBH energy density and
\be
\label{dd}
M_*(z)=N_*(z)\cdot  \mpbh\simeq \fpbh^2\left(\frac{2600}{1+z}\right)^2\mpbh
\ee
is the typical mass of halos collapsing at redshift $z$, see also~\cite{hutsi}. In the halo model framework \cite{coo}, the correlation function in the non-linear regime may also be written as~\cite{MS,sj}
\be
\label{f}
\xi(x,z)=\frac{1}{\overline{\rho}^2_\text{\tiny DM}}\int\d M\,  \frac{\d n(M,z)}{d M} \,M^2\, \lambda_{\tiny M}(x,z),
\ee
where 
\begin{equation}
  \lambda_{\tiny M}(x,z)=\int \d^3 s\, \rhopbh(s,M,z)\rhopbh(|{\vec s}+{\vec x}|,M,z)\simeq \frac{1.22}{4\pi R_{\text{\tiny vir}}^3}\left(\frac{x}{R_{\text{\tiny vir}}}\right)^{-9/5},
\end{equation}
in terms of the average density profile of a halo of mass $M$~\cite{sj}
\be
\rhopbh(x,M,z)=\left(\frac{3}{5\cdot 4\pi R_{\text{\tiny vir}}^3}\right)\left(\frac{x}{R_{\text{\tiny vir}}}\right)^{-12/5},
\ee
and the virial radius $R_{\text{\tiny vir}}$ defined through
\begin{equation}
\label{Rvir}
  R_{\text{\tiny vir}}^3=\left(\frac{3M}{4\pi\cdot 200\,\overline{\rho}_{\text{\tiny PBH}}}\right)
\end{equation}
assuming an average overdensity $\sim 200$ within each virialized halo. 
It is easy to show that the mass integral in Eq.~(\ref{f}) gets its largest contribution from halo masses around $(11/10)M_*(z)$. In other words, halos with the characteristic mass $\sim M_*(z)$ give the largest contribution to the correlation function at a given redshift $z$.

Notice that Ref.~\cite{Trashorras:2020mwn} reported a steeper scaling of the PBH profile, with $\rho_\PBH(r)\sim r^{-2.8}$ between $(10^{-3} \div 10) \, {\rm pc}$. One should notice though that their simulations
involve the evolution of a single cluster (as in Ref. \cite{j1}) and assume a clustering of about $10^3$ PBHs with $f_\PBH=1$ already at redshift $z\sim 10^3$ and at kpc comoving scales, while  at that redshift the Press-Schechter theory, tested  also in  Ref.~\cite{inman}, predicts that the typical halo has only a few PBHs.
	
Of course our results are limited in various aspects. First, we have employed the stable clustering hypothesis which is only valid prior to a cosmological constant- or dark energy-dominated period. Secondly, binaries tend to sink towards the center of halos since they are heavier than single PBHs, and eventual binary-PBHs interactions will heat up the core, possibly modifying its shape. Thirdly, other many phenomena should occur at the very small scales, including encounters with massive PBHs in the core which cause the lighter PBHs to be ejected and form their own, albeit shallower profile \cite{Trashorras:2020mwn}.

	\section{Evaporation and the halo survival time}
	\noindent
	Our analysis of the PBH correlation function do not take into account the  evaporation of PBHs from the edges of the  cluster. In this section,
	we show that this effect is likely not relevant owing to the competing accretion of smaller halos into bigger ones, at least for 
	the  interesting case $f_\PBH = 1$ which we shall focus on hereafter.
	
	The formation redshift of a cluster of $N=M/\mpbh$ PBHs can be estimated from Eq.~(\ref{dd}),
	\be
	1+z_\text{\tiny form} = \frac{2600}{\sqrt{N}}.
	\ee
	Random encounters can give a PBH enough energy to escape from the halo. 
	The evaporation time of a system of $N=M/\mpbh$ PBHs   clustered in a region of size $R$ and subject to the gravitational force is given by~\cite{binn}
	\begin{equation}
	t_{\text{\tiny{ev}}}
	\simeq14 \frac{N}{\log N} \frac{R}{v},
	\end{equation}
	where $ v \simeq \sqrt{GN M_\PBH/R}$. Therefore, for the typical cluster virialization radius $R_{\text{\tiny vir}}$,
	\begin{equation}
	t_{\text{\tiny{ev}}} \simeq \frac{8 \cdot  10^{20} {\rm s}}{\log N} \left( \frac{N}{100} \right)^{1/2} \left(  \frac{\mpbh}{ 20 M_\odot/h}\right)^{-1/2}  \left( \frac{R_{\text{\tiny vir}}}{{\rm kpc}/h} \right)^{3/2}.
	\end{equation}
	We must now assess whether a given halo has enough time to evaporate before being included in a bigger halo. The survival time of a given halo of mass $M$ can be computed by resorting again to the Press-Schechter formalism. 
	Following Ref.~\cite{lc}, we define a time dependent threshold for collapse as $\omega (z) \equiv \delta_c / a = \delta_c (1+z)$ (we use again a matter-dominated period) and a time independent variance as
	\begin{equation}
	S (R) = (1+z)^2\int \frac{ \d k}{k} \Delta^2_\text{\tiny L}(k,z) W^2 (k R),
	\end{equation}
	where $W (k R)$ is the Fourier transform of a top-hat window function.
	For the linear power spectrum at hand, one can take advantage of the relation
	\begin{equation}
	\frac{\omega}{\sqrt{S}} = \lp \frac{M}{M_*} \rp^{1/2} = 2  \cdot 10^{-4}\lp \frac{ M}{M_\PBH}\rp ^{1/2}  \cdot 1.68  \lp 1+z \rp,
	\end{equation}
	from which we read off
	\begin{equation}
	S (M)= 2.5 \cdot 10^7 \lp \frac{ M}{M_\PBH}\rp ^{-1} = 2.5 \cdot 10^7 N^{-1}.
	\end{equation}
	\begin{figure}
		\centering
		\includegraphics[width= .7\linewidth]{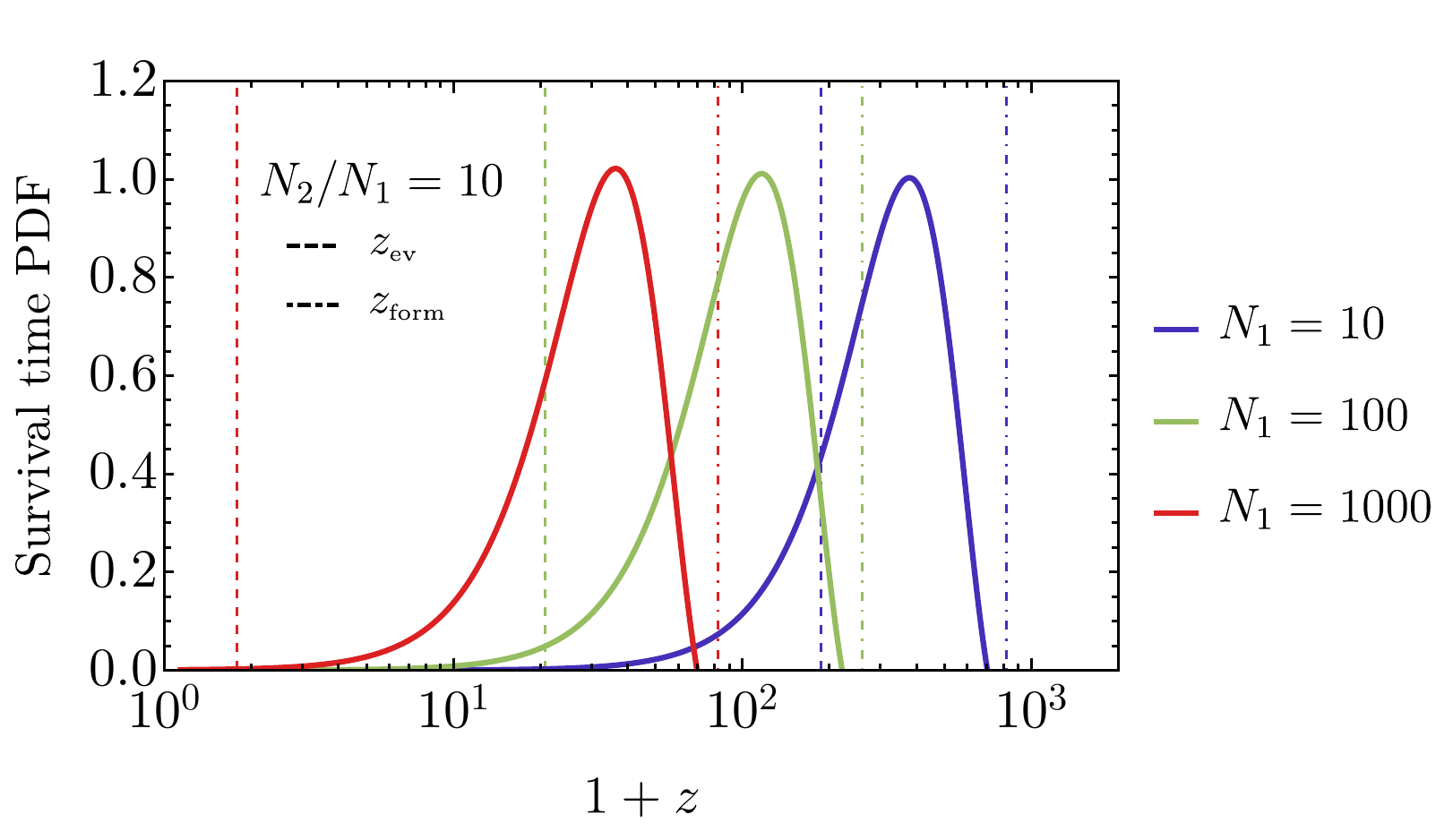}
		\caption{Probability distribution for the survival time of a halo with $N_1$ PBHs at $z=z_\text{form}$ until it is incorporated into a bigger halo with $N_2= 10 N_1$ PBHs. Dashed (dot-dashed) vertical lines indicate the characteristic evaporation (formation) redshift of the cluster of $N_1$ objects.}
		\label{fig4}
	\end{figure}
	\noindent
The probability that a halo of mass $M_1$ formed at redshift $z_\text{\tiny form} (M_1)$ (corresponding to a variance $S_1$ and time $\omega_1$) is incorporated in a bigger halo of mass $M_2$ at a subsequent redshift (corresponding to $S_2$ and $\omega_2$) is given by~\cite{lc}
	\begin{align}
	g (S_2, \omega_2 | S_1, \omega_1) \d \omega_2 &= \sqrt{\frac{2}{\pi}} \frac{1}{\omega_1} \sqrt{\frac{S_1}{S_2(S_1-S_2)}} \exp \llp \frac{2 \omega_2 (\omega_1 - \omega_2)}{S_1} \rrp  \left\{ \frac{-S_2 (\omega_1 - 2 \omega_2) - S_1 (\omega_1 - \omega_2)}{S_1  e^{X^2} }\right. \nonumber \\ 
	& \left. + \sqrt{\frac{\pi}{2}} \sqrt{\frac{S_2(S_1-S_2)}{S_1}} \llp 1 - \frac{(\omega_1 - 2 \omega_2)^2}{S_1}\rrp  \llp 1-\text{erf}(-X) \rrp 
	\right\} \d \omega_2,
	\end{align}
	where $\omega_2 < \omega_1$, $S_2 < S_1$ and
	\be
	X = \frac{S_2(\omega_2-2 \omega_1) + S_1 \omega_2}{\sqrt{2 S_1 S_2 (S_1 - S_2) }}.
	\ee
	Since clustering is hierarchical, each halo has a certain survival time and, therefore, a given probability to be absorbed by a bigger halo formed at a later redshift. 
	As we will see, the evaporation time of PBH halos is typically larger than their survival time, which implies that PBH halos are stable against evaporation.
	We illustrate this phenomenon in Fig.~\ref{fig4}, which shows the probability that a halo containing $N_1$ PBHs  at redshift $z_\text{\tiny form}$ is incorporated into a bigger halo containing $N_2 = 10 N_1$ PBHs at redshift $z<z_\text{\tiny form}$. In all cases, the peak of the distribution occurs before the evaporation redshift $z_\text{\tiny ev}$ shown as the vertical dashed line.
	Notice that the characteristic survival redshift of the progenitor halo of mass $N_1 M_\PBH$ broadly agrees with the formation time of the descendant halo of mass $N_2M_\PBH$ (for $N_1=10$ and 100), which reflects the consistency of the excursion set approach used here.
	
	We conclude that the correlation function is not altered by evaporation at least in the interesting case in which $f_\PBH = 1$.
	For smaller values of $f_\PBH$, we expect the dynamics to be more complicated due to the presence of an additional component of DM, yet also clustering to be much less relevant.

	\section{Impact on  merger rates }
	\noindent
	The next question we want to address is the impact of PBH clustering onto the merger rates of PBHs. We shall distinguish between the impact of clustering on the merger rate of PBH binaries formed in the early and in the late Universe.
	\subsection{Merger rate from PBH binaries formed in the early Universe}
	\noindent
	The formation of PBH binaries in the early Universe typically occurs deep in the radiation epoch \cite{revPBH} when fluctuations in the PBH number counts are still Poissonian. Since PBHs are not significantly clustered at their formation epoch \cite{cl1,cl2,cl3,cl4}, we conclude that the formation of PBH binaries in the early universe is not altered by clustering. However if the PBH binaries formed in the early Universe end up in highly clustered PBH regions, there is a greater chance they are perturbed by close encounters with another PBH. This three-body interaction changes the semi-major axis of the binary and moves the  eccentricity $e$ away from  its initial  high value  $e\simeq   1$ \cite{raidal,j1}. 
	Since the  coalescence time due to the emission of gravitational waves is 
	\be
	t_{\text{\tiny GW}}\simeq \frac{3}{170}\frac{1}{(G \mpbh)}a^4(1-e^2)^{7/2},
	\ee
	clustering thus enhances departures from the initially high eccentricity regime and, consequently, increase significantly the coalescence time.
        It has been argued therefore that the early Universe merger rates of PBHs in the LIGO/Virgo mass range are suppressed for $\fpbh\simeq 1$, when clustering is stronger, thus providing rates much smaller than those required by the observed events \cite{j1,raidal}. However, one should recall two effects that tend to  reduce  the frequency of binary-PBH encounters ~\cite{Vaskonen:2019jpv}.  First, the disruption of binaries residing in DM  halos decreases  when smaller halos merge into bigger halos \cite{k} and, second, halos  expand due to the heating provided by binary-PBH collisions. 
	
	To be conservative and since the early Universe merger rate for unperturbed PBH binaries is above the LIGO/Virgo detection band, 
	we focus here on the effects which may increase the disruption of PBH binaries and decrease the merger rate. First of all, binaries, being heavier  
	than a single PBH, sink towards the halo center and, secondly, the halo core where velocities are peaked is subject to a gravothermal instability (arising from the negative heat capacity of self-gravitating systems \cite{ldw}) triggered by PBH evaporation from the cluster and, therefore, may collapse.
	When the core contracts, the central density increases, leading to more frequent binary-PBH encounters which may eventually halt the collapse.  
	Following Ref. \cite{Vaskonen:2019jpv}, we assume that, if the halo core is unstable, all PBH binaries are perturbed on a timescale smaller than the age of the Universe at a given redshift. The gravothermal instability timescale is given by \cite{Quinlan:1996bw}
	\be
	t_\text{\tiny GI} = \frac{v^3(r)}{G^2M_\PBH \rho_\PBH (r) \log (M (<r)/ M_\PBH)},
	\ee
	where we adopt the PBH density profile appropriate to a halo mass $M_h$ given in the previous section,
	\be
	\rho_\PBH(r) \simeq \frac{3 M_h}{20 \pi}  R_\text{\tiny vir}^{-3/5}  r^{-12/5}.
	\label{ll}
	\ee
	The normalisation constant is found by imposing that the halo is composed of PBHs only, i.e. $\fpbh=1$ as assumed throughout this section.
	Furthermore, the virial radius can be estimated by imposing that the mean density within the radius $R_\text{\tiny vir}$ is $\rho_\PBH(<R_\text{\tiny vir}) = 200 M_\PBH \bar n_\PBH$.
	The characteristic relative velocity at a given radius $r$ is then given by
	\begin{equation}
	v (r) = \sqrt{\frac{G M (<r)}{r}} \simeq \sqrt{G M_h R_\text{\tiny vir}^{-3/5} }  r^{-1/5}.
	\end{equation}
	Requesting the gravothermal timescale to be less than the Hubble time and replacing the halo mass $M_h$ with the characteristic value $M_*(z)$, we find the critical radius below which the instability is rapid enough to occur within a Hubble time so that binaries are perturbed. At $z=0$ for instance, the critical radius is $\sim 3\cdot 10^{-3}$ kpc/$h$ corresponding to a a critical number of PBHs $N_\text{\tiny c}\sim 4.6 \cdot 10^4$.

        Next, we calculate the fraction of initial PBH binaries contained in gravothermally unstable cores.
        For an initially Poisson distribution, the probability of finding a PBH within a halo made up of $N$ PBHs at redshift $z$ is \cite{epst} (see also \cite{rks98})
	\be
	p_N(z)\propto N^{-1/2} e^{-N/N_*(z)}.
	\ee
	Therefore, the  probability of having a binary in a halo of $N$ PBHs is approximately proportional
	to $p_N$, while the probability of  finding a PBH
	in a subhalo of $N$ PBHs embedded in a parent halo of $N' > N$ PBHs is proportional to 
	$p_N\cdot p_{N'}$  \cite{Vaskonen:2019jpv}. The fraction of unperturbed binaries at the present time is thus bounded from below by
	\be
	P_\text{\tiny np}\gsim 1-\sum_{N=3}^{N_\text{\tiny c}}\overline{p}_N(z^\text{\tiny c}_\text{\tiny form})-\sum_{N'>N_\text{\tiny c}}\left[
	\sum_{N=3}^{N_\text{\tiny c}}\widetilde{p}_N(z^\text{\tiny c}_\text{\tiny form})\right]\overline{p}_{N'}(z^\text{\tiny c}_\text{\tiny form})\simeq 10^{-2},
	\ee
	where $z^\text{\tiny c}_\text{\tiny form}$ is the formation time of the halo with $N_\text{\tiny c}$ PBHs and
	\be
	\sum_{N\geq 2} \overline{p}_N=1\,\,\,\,{\rm and}\,\,\,\, \sum_{N= 2}^{N'} \widetilde{p}_N=1.
	\ee
	Hence, the corresponding early Universe merger rate of unperturbed binaries is given by ${\cal V}^\text{\tiny EU}_\text{\tiny np}\cdot P_\text{\tiny np}$ where \cite{raidal}, see also \cite{noi3},
	\begin{align}
	\mathcal{V}^\text{\tiny EU}_\text{\tiny np}  &\simeq 7.5 \cdot 10^{4} \,\lp \frac{M_\PBH}{20 M_\odot/h} \rp^{-32/37} {\rm Gpc^{-3}} {\rm yr^{-1}}.
	\end{align}
	Consequently, ${\cal V}^\text{\tiny EU}_\text{\tiny np}\cdot P_\text{\tiny np}$ remains above the LIGO/Virgo detection band for all  the PBH masses detectable by the collaboration\footnote{The detection band is  $\sim (10\div 10^2)\, {\rm Gpc^{-3}} {\rm yr^{-1}}$ for PBH masses $\sim  (20\div 30)\, M_\odot$  and has an upper bound of $\sim 5\cdot 10^3\, {\rm Gpc^{-3}} {\rm yr^{-1}}$ for PBH masses $\sim M_\odot$ \cite{LIGOScientific:2018mvr}.}. Furthermore, one should bear in mind that there are other contributions to the merger rates. First,  there are perturbed PBH binaries whose binary parameters still allow for the coalescence time to be comparable to the current age of the Universe; secondly,  not all the binaries end up  inside halos. For $\fpbh\simeq 1$ and using the probability $p_N$, one can easily estimate that $\sim 10^{-3}$ PBHs are  not in clusters  (for definiteness we consider halos with at least ten PBHs as in Ref.~\cite{k}), leading to a merger rate $\sim 75 (h M_\PBH/20 M_\odot)^{-\frac{32}{37}} {\rm Gpc^{-3}} {\rm yr^{-1}} $, which is at best close to the upper bound given by LIGO/Virgo  for PBH masses of order $\sim M_\odot$.
	
	We conclude  that the  merger rate of PBH binaries formed in the early Universe in the presence of clustering
	is likely to be above the  LIGO/Virgo detection band.

	\subsection{Merger rate from PBH binaries formed in the late Universe}
	\noindent
	We now discuss the impact of clustering onto the late time PBH binary merger rate for $f_\PBH=1$.  If a PBH moving at a given velocity $v$ passes close to another PBH, the cross section for binary formation at late epochs is given by~\cite{Bird:2016dcv}
	\begin{equation}
	\sigma_\text{\tiny bin} \simeq \lp\frac{85\pi }{3} \rp ^{2/7} \frac{\pi \lp2 G M_\PBH \rp^2}{v^{18/7}}
	\end{equation}
        in the Newtonian limit. 
	Once it has formed, such a binary can merge within the age of the Universe. For a halo of mass $M_h$, the merger rate can be computed as~\cite{Bird:2016dcv}
	\begin{equation}
	R_h(M_h) =  2 \pi\int _{0} ^{R_\text{\tiny vir}} \d r  \, r^2 \lp \frac{\rho_\PBH(r)}{M_\PBH}\rp ^2 \langle \sigma_\text{\tiny bin} v\rangle,
	\end{equation}
	where the brackets stand for the usual thermally averaged cross-section, i.e. the mean of the combination $\sigma_\text{\tiny bin} v$ with velocities drawn from a Maxwell-Boltzmann distribution.
	For simplicity, we assume that the PBH  density profile  is constant within the core of size $r_s$ (to be determined below), while it scales like $\rho_\PBH(r)\sim r^{-12/5}$ for $r > r_s$.
	The resulting present-day merger rate reads
	\begin{align}
	\mathcal{R}_h(M_h) & \simeq
	22   \lp G M_h^{4/5}  M_\PBH^{1/5} \bar{n}_\PBH^{1/5}   \rp^{17/14}   r_s^{-52/35}.
	\end{align}
	The total merger rate is obtained by convolving the merger rate per halo $\mathcal{R}_h$ with the halo mass function $\d n/\d M_h$ derived in Eq.~\eqref{halo function} using the Press-Schechter formalism,
	\be
	\mathcal{V}_\text{\tiny LU} = \int \d M_h \frac{\d n}{\d M_h} \mathcal{R}_h(M_h)\simeq 1.5 \cdot 10^3 \,G^{17/14} (M_\PBH \bar{n}_\PBH)^{73/42}  M_*^{-11/21} {\cal R}_{\text{\tiny cl}}^{52/35},
	\ee
	where the dimensionless ``cluster factor''
	\be
	{\cal R}_{\text{\tiny cl}}=\frac{R_*}{r_s}
	\ee
	is expressed  in  terms of the characteristic scale $R_*\simeq 9 \lp  h M_\PBH/20 M_\odot\rp^{1/3}$ kpc$/h$  identified with the virial radius of an  halo of mass $M_*$. Inserting the value for the mean present halo mass $M_* = 6.8 \cdot 10^6 M_\PBH$, we arrive at
	\begin{align}
	\mathcal{V}_\text{\tiny LU}  &\simeq 10^{-4}   \lp \frac{M_\PBH}{20 M_\odot/h} \rp^{-11/21} {\cal R}_{\text{\tiny cl}}^{52/35} {\rm Gpc^{-3}} {\rm yr^{-1}}.
	\end{align}
	We infer that the current LIGO/Virgo merger rate detection band $(10\div 10^2) $ ${\rm Gpc^{-3}} {\rm yr^{-1}} $ for PBH masses around 20 $M_\odot$ is matched with ${\cal R}_{\text{\tiny cl}}\simeq (10^3\div 10^4)$ or, equivalently, with
	$r_s\simeq (10^{-3}\div10^{-2}) \lp  h M_\PBH/20 M_\odot\rp^{1/3}$ kpc$/h$. This estimate assumes that the power-law shape of the PBH correlation function remains valid down to small scales and until late times. We expect the value of $r_s$ to depend on the details of the  dynamical processes responsible for core collapse.  Initially,  the core contracts in order to conserve energy as PBHs evaporate from the high tail of the velocity distribution. As the collapse proceeds, the core becomes hotter. However, when the  core is small, binaries can form and harden, which could possibly stop (and even reverse) the collapse and set a minimum radius $r_s$.
	
	The latter can be roughly estimated bby requiring the gravothermal instability timescale to be smaller than the age of the Universe. For the characteristic mean halo mass $M_*$ giving the dominant contribution to the merger rate, this yields a minimum radius of order $r_s \simeq 3 \cdot 10^{-3} {\rm kpc}/h$ and a cluster factor of ${\cal R}_{\text{\tiny cl}} \simeq 3\cdot 10^3$. \footnote{A similar value is obtained by assuming that a halo forms a core of radius $r_s$ due to the gravitational PBH interactions equalising the kinetic energies \cite{sh}. Imposing the relaxation time of the core  to be smaller than the age of the Universe, one obtains ${\cal R}_{\text{\tiny cl}} \simeq 10^{2}$, which would imply a late Universe merger rate still below the LIGO/Virgo detection band.}
	
	We conclude that, for large $f_\PBH$, clustering may help increasing the late time merger rate  so that it is visible in the LIGO/Virgo band.
        However, PBH clustering decreases the early Universe merger rate, yet not enough for it to be consistent with current LIGO/Virgo data.
	A deeper understanding of the minimum radius $r_s$ along with detailed N-body simulations (down to low redshift) is, of course, required before drawing any firm conclusion.

	\section{Conclusions}
	\setcounter{section}{5}
	\noindent
	The clustering of PBHs is a crucial ingredient which may significantly affect the merger rates of coalescing binaries and, consequently, the gravitational wave signal measured by the LIGO/Virgo collaboration. Furthermore, it is relevant for the interpretation of the constraints on PBH abundance and masses. We have provided some analytical insights into PBH clustering assuming that PBHs are initially Poisson distributed. We have also investigated its impact on the early and late Universe merger rates. 
Our findings indicate that
	\begin{itemize}
	\item for a small fraction of PBH contribution to the DM, PBH clustering does not affect the standard calculation of the merger rates;  
	
	\item the evaporation phenomenon is not likely to change the clustering properties of PBHs;
	
	\item the early Universe merger rate is decreased in the presence of clustering for large $f_\PBH$, yet still falling above the LIGO/Virgo detection band;
	
	\item the late Universe merger rate is increased and can fall within the detection band.
		\end{itemize}
	Clustering might also be relevant for PBH spins. While the gravitational collapse of a spherical overdensity during radiation domination generates nearly spinless PBHs \cite{ds,mgn}, they may acquire a large spin in the presence of accretion \cite{ps}.  If PBHs cluster, then a large number of them can merge to form binaries in the late time Universe,
and it is expected that the spin of the resulting PBHs from each merger event has a non-zero value even when no accretion is present. 

Our results requires various refinements. For instance,  the inclusion of the late time effect of the cosmological constant and a thorough numerical investigation (along the lines of \cite{inman}) to confirm  the scaling law  at  small-scales.   It also remains to be seen if an universe in which PBHs make up all the DM is compatible with observations. In this regard, we can think of a few relevant questions:
	
	\begin{enumerate}
	\item Is the formation of the galaxies in clusters consistent with $\fpbh=1$? The characteristic PBH halo mass today is $\sim 10^7(\mpbh/M_\odot)\,M_\odot$ which is much smaller than the typical galaxy mass, e.g. the Milky Way.
	\item Are the bounds on PBHs from the Lyman-$\alpha$ forest strengthened with clustering?
	
	\item What is the impact of clustering on the idea of distinguishing astrophysical from primordial BHs using the cross-correlation with the galaxies?
	
	\item What is the impact of clustering on the mixed merger rate of PBHs with astrophysical BHs?
	\end{enumerate}
	We plan to return to these issues in the future.

	\begin{center}
		{\bf  Acknowledgments}
	\end{center}
	\noindent
	We especially thank D.~Inman and Y.~Ali-Ha\"{i}moud for providing us with extra data from the simulations of Ref.~\cite{inman}, and H.~Veerm\"{a}e and M. Raidal for informations about Fig.~7 of Ref.~\cite{raidal}.
	We are indebted with M. Biagetti, E. Branchini and P. Pani  for stimulating discussions and for useful comments on the draft. 
	V.DL., G.F. and A.R. are supported by the Swiss National Science Foundation 
	(SNSF), project {\sl The Non-Gaussian Universe and Cosmological Symmetries}, project number: 200020-178787.
        V.D.  acknowledges support by the Israel Science Foundation (grant no. 1395/16). 
	
	\appendix

	\section{The clustering of the extra non-relativistic DM component}
\begin{figure}[t!]
		\centering
		\includegraphics[width= .7\linewidth]{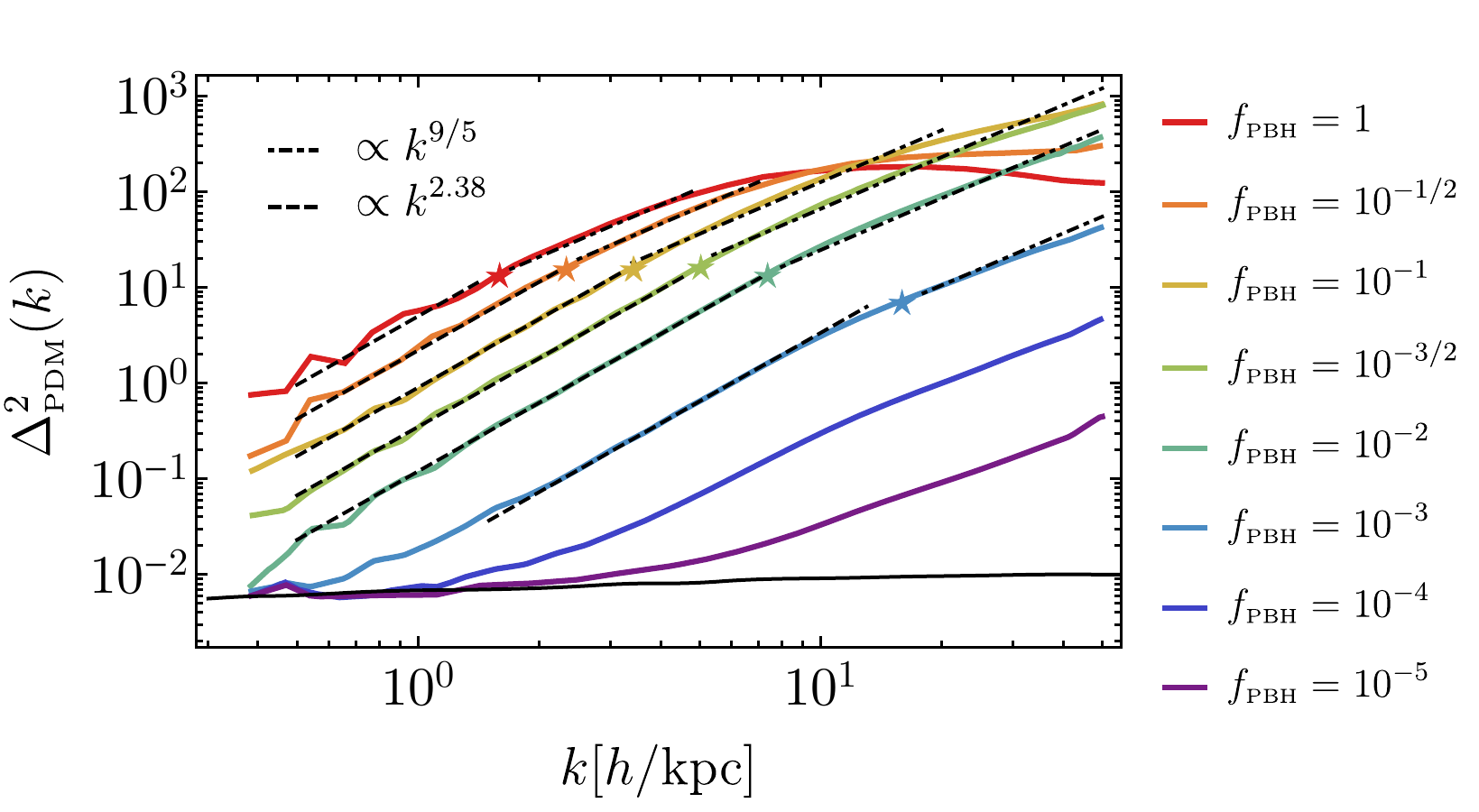}
		\caption{PDM power spectrum as a function of the momentum scale $k$ for a fixed $f_\PBH$ from \cite{inman}.
			The plotted stars indicate the inverse of the characteristic PDM structure size for each value of $f_\PBH$ computed by equating the linear PDM smoothed density contrast to the quasi-linear threshold $5.85$, see Eq.~\eqref{nlcscale}. The dashed and dot-dashed black lines indicate the analytical fits in the two regimes described in the text. The black solid line indicates the value of the adiabatic perturbations found in linear theory matched by the simulation results for low enough PBH abundances \cite{inman}.}
		\label{fig.PDM}
	\end{figure}

	\renewcommand{\theequation}{A.\arabic{equation}}
		\noindent
If the PBHs do not comprise all the DM, there must be another non-relativistic component, which was dubbed in Ref. \cite{inman} Particle Dark Matter (PDM).
	In this appendix we offer some considerations about its clustering. The  PDM has on large scales a small adiabatic component characterized by a power spectrum which is almost flat and normalised to the CMB anisotropy. We neglect it from now on. Instead, the PDM falls into the potential wells of the PBHs already at the linear level enhancing the perturbations (see Fig.~\ref{fig.PDM}), such that \cite{inman}
	\begin{align}
	\Delta_\text{\tiny PDM}^{2} (z,k) & \simeq  \lp \frac{3}{2} \frac{1+z_\text{\tiny eq}}{1+z} \rp^2 \Delta^2_{\text{\tiny i}}(k). 
	\end{align}
	The typical PBH halo contains at a redshift $z$ a number of PBHs given by $\sim f^2_\PBH(2600/1+z)^2$. Taking  $z=100$ to compare to the findings of Ref. \cite{inman}, this gives $\sim (26 f_\PBH)^2$, which is larger than unity for 
	$f_\PBH\gsim 0.04$.  Therefore for  $f_\PBH\lsim 0.04$, PBHs are sparse and they have a Poisson distribution. PDM falls into their potential wells with an average profile at the linear level given by  
	\be
	\overline{\delta\rho}_\text{\tiny PDM}({\vec x},z)\simeq \frac{\langle \delta\rho_\text{\tiny PDM}({\vec x},z)|\delta\rho_\text{\tiny PBH}(0,z) \rangle}{\langle \delta\rho^2_\text{\tiny PBH}(0,z)\rangle}\
	\delta\rho_\text{\tiny PBH}(0,z)\sim \xi_\text{\tiny L}(x,z)\sim x^{-3}.
	\ee
	At the linear level PDM is therefore peaked around PBHs with a profile which decays like $x^{-3}$. If  the initial density profile of the PDM  halo is a power law $\sim x^{-\epsilon_\text{\tiny PDM}}$ in radius and the initial progenitor PDM is  related to peaks in the initial density field, then following Refs.~\cite{cr1,cr2}, one can argue that the collapsed PDM halo
	has a power law profile $\sim x^{-3\epsilon_\text{\tiny PDM}/(1+\epsilon_\text{\tiny PDM})}=x^{-9/4}$ for $\epsilon_\text{\tiny PDM}=3$.  However, once the clustering of the PDM becomes efficient, the presence of the PBH becomes irrelevant as it acts only as an initial catalyzer.  It is not clear therefore what exact value of $\epsilon_\text{\tiny PDM}$ is the relevant one. This is because in peak theory a crucial role is also played by the second derivative of the peak profile, which goes like
	$\sim x^{-5}$ \cite{bbks}. Thus, setting $\epsilon_\text{\tiny PDM}=4$ should give an indication of the peak profile \cite{sj}. This in turn would give a final PDM profile of $\sim x^{-12/5}$. Similarly to the PBH clustering,  one therefore expects that the PDM power spectrum in the non-linear regime will go like $\sim k^{2\cdot12/5-3}=k^{9/5}$, while in the quasi-linear and linear regime it should go like the profile itself, that is $\sim k^{12/5} \simeq k^{2.38}$. This expectation is matched as seen in Fig. \ref{fig.PDM}. The transition between the different power laws happens when the PDM becomes non-linear and is indicated in the figure by stars.  To estimate the corresponding scales we can proceed as follows. 
	The linear PDM variance smoothed at a given scale $R$ is given by 
	\begin{equation}\label{eqsigma}
	\sigma_\text{\tiny PDM}^2(R) = \int \frac{ \d k}{k} \Delta^2_\text{\tiny PDM}(k) W^2 (k R),
	\end{equation}
	where $W(kR)$ is the Fourier transform of the window function, which we take to be a Gaussian function. We find
	\begin{equation}
	\sigma_\text{\tiny PDM} (R) =  f_\PBH \lp \frac{1+z_\text{\tiny eq}}{1+z} \rp\frac{\pi^{1/4}}{2} (k_* R)^{-3/2}.
	\end{equation}
	Imposing $\sigma_\text{\tiny PDM}(R_\text{\tiny PDM}^\text{\tiny NL})\simeq 5.85$ we get
	\begin{equation}\label{nlcscale}
	R_\text{\tiny PDM}^\text{\tiny NL}= 5.4 \cdot 10^{-2}  \,  f_\PBH^{1/3} \lp \frac{1+z}{1+z_\text{\tiny eq}} \rp^{-2/3}\lp \frac{M_\PBH}{20 M_\odot/h}\rp^{1/3}\, {\rm kpc}/h.
	\end{equation}
	For $f_\PBH\gsim 0.04$, the typical  PBHs halo contains   more than one PBH, and thus PBHs cluster.  From our previous results, the expected PBH density profile goes  as $\rho_\text{\tiny PBH}^\text{\tiny NL} \sim r^{-12/5}$. Following Ref.~\cite{Bertschinger:1985pd}, one can define the  turn-around radius $r_\text{\tiny ta}$   where the PDM decouples from the background expansion, overcoming the outward inertia, and collapses onto the PBHs. 
	The total PBH mass within such physical radius is  therefore
	\be
	\frac{1}{2}M_\PBH(< r_\text{\tiny ta})= \frac{1}{2} \frac{4 \pi}{3} \rho_\text{\tiny PBH}^{\text{\tiny{NL}}}(r_\text{\tiny ta}) r_\text{\tiny ta}^3 = \frac{4\pi}{3} \overline{\rho}\,r_\text{\tiny ta}^3.
	\ee
	A simple estimate of the PDM density profile surrounding  the PBHs can be obtained by assuming that the PDM is  frozen in at turn-around with their density matching the 
	background density at that time. This is essentially the circular orbit model of Ref. \cite{rg}, in which mass shells are placed on circular orbits with energy equalling that at turn-around.
	One finds
	\be
	r_\text{\tiny ta}^{-12/5} \sim H^2 \sim t^{-2}, 
	\ee
	where $H$ is the Hubble rate. In this way one obtains 
	in matter-domination  a PDM profile as
	\be
	\rho_\text{\tiny PDM} = \frac{\overline{\rho}_\text{\tiny eq}}{2} \lp \frac{a}{a_\text{\tiny eq}}\rp^{-3} =  \frac{\overline{\rho}_\text{\tiny eq}}{2}
	\lp \frac{t}{t_\text{\tiny eq}}\rp^{-2} 
	\ee
	that is 
	\be
	\rho^{\text{\tiny NL}}_\text{\tiny PDM} (r) \sim r^{-12/5}\,\,\, \text{or} \,\,\, \xi^{\text{\tiny NL}}_\text{\tiny PDM} (r) \sim r^{-9/5} \,\,\, \text{and} \,\,\, \Delta^\text{2\tiny NL}_\text{\tiny PDM} (k) \sim k^{9/5},
	\ee
	which is again in good agreement with the numerical results of Ref. \cite{inman} plotted in Fig. \ref{fig.PDM}\footnote{Notice that for large values of $f_\PBH$ the PDM power spectrum switches off at large momenta, most probably because the PDM  halo profiles around a single PBH overlap in the presence of many PBHs, which tend to generate an overall flat profile \cite{inman}.}.

\section{Describing  BHs in an expanding universe}
\noindent
\renewcommand{\theequation}{B.\arabic{equation}}
Since in this paper we touch upon the limits imposed by the LIGO/Virgo collaboration, we offer some considerations about a recent claimed done in 
Ref.~\cite{Boehm:2020jwd} where it was proposed that the LIGO/Virgo bounds are  relaxed by asserting that  the compact objects seen by the LIGO/Virgo collaboration should be in fact  described by  BHs  in an expanding  universe and therefore characterized by a growing mass $m a(t)$, where $a(t)$ is the scale factor. 

We argue here that a consistent and standard description of a constant mass  BH in an expanding universe is possible and therefore  that the LIGO/Virgo constraints are  not relaxed.

Let us first summarise  the argument in Ref. \cite{Boehm:2020jwd}. 
The starting point is the assumption that a BH in a Friedmann-Robertson-Walker (FRW) expanding  universe is described by the metric
\begin{equation}
	\d s^2 = f(R) \lp 1- \frac{H^2 R^2}{f^2(R)} \rp \d t^2 
	+ \frac{2 H R}{f(R)} \d t \d R 
	- \frac{\d R^2}{f(R)}
	- R^2 \lp \d \theta^2 + \sin^2 \theta \d \phi^2 \rp,
\end{equation}
where $f(R) = 1 - 2  G m a(t)/R(t)$ and $R(t) = a(t) r$ is the physical radial coordinate. Let us stress here that $m$ is a constant parameter which is identified with the observable mass today. It is clear that, in such a metric, the would be ``Schwarzschild horizon" is growing proportional to the scale factor, thus being comoving. 
Following the definition of the quasi-local Misner-Sharp mass \cite{Misner:1964je},
 one can read the mass from the $g_{00}$ component of the metric as
\be
m_\text{\tiny MS} = m a(t) + \frac{H^2 R^3}{2G f(R)},
\ee
which, few e-fold after the BH formation,  is dominated by the first term due to the time evolution (using $H^2 \sim a^{-4}$ and $R\sim a$ valid in a radiation-dominated universe).

An object with a final mass of the order of few tens of solar masses, as the one observed by the LIGO/Virgo collaboration, would have possessed a smaller mass $m_\text{\tiny MS} \simeq m a(t) $ at higher redshifts. Imposing the condition of decoupling from the Hubble flow is satisfied, which means the gravitational attractive force is larger that the expansion force, requires the PBHs to have very small separations. This in turn implies that the merger time due to the subsequent emission of GWs is much smaller than the age of the universe for binaries which decouple before structure formation, thus avoiding the LIGO/Virgo bounds on early universe binaries.

However, it appears that the metric giving rise to a comoving horizon are often ill defined, see for example Ref.~\cite{Faraoni:2007es}. 
Also, in order to have such comoving BH solutions, a very specific cosmic fluid dominating the universe energy budget needs to be 
added to the dynamics.  Last, but not least, such an  exotic object would behave on cosmological scales as a strongly accreting DM component which may prevent them from being a good DM candidate.

In the following we will show that there exists a fully consistent description of a BH  in an expanding FRW universe whose dynamics and properties are basically not  modified with respect to the standard description where the expansion of the universe is neglected.

General Relativity solutions describing a BH in a FRW universe can have different properties. A particularly interesting class of solutions is provided by the McVittie metric in terms of the scale factor $a(t)$ and the BH mass $m$ \cite{Kaloper:2010ec}
\begin{equation}
\d s^2 = - \lp \frac{1-\mu}{1+\mu} \rp^2 \d t^2 +\lp 1+ \mu \rp^4 a^2(t) \d \vec x ^2,
\end{equation}
where
\begin{equation}
\mu = \frac{G m}{2 a(t) | \vec x |}.
\end{equation}
Notice that the McVittie solution is found by explicitly assuming a ``non-accreting" condition for the BH mass, that is
$
m = {\rm const}.
$
In particular, this condition implies the metric component $g^{0r}$ to be vanishing. It is therefore easy to show, using the Einstein's equations,  that also the matter energy-momentum tensor should have $T^{0r} = 0$. Therefore, the McVittie solution is the only possible solution where a single perfect fluid dominating the energy density of the universe is considered. \footnote{Here we make use of the definition of the perfect fluid as the one described by an energy-momentum tensor which can be set in the diagonal form $T_{\mu \nu} = {\rm diag} \lp \rho, P,P,P \rp$.}

The aformentioned assumption needs to be relaxed in order to allow for GR solutions with accreting BHs, see for example Ref.~\cite{Faraoni:2007es}, which are however typically plagued by tachyonic instabilities or superluminal speed of the cosmic fluid.

If one expands the metric in the limit of small $\mu$ (large distances) one finds an FRW universe in the Newtonian gauge 
in the presence of perturbations given by the BH Newton potential (with constant mass and scaling like $1/r$, where $r \simeq a(t) |\vec x|$ is the physical distance) as 
\begin{equation}
\d s^2 = - \lp 1- \frac{2 G m}{a(t) | \vec x |} \rp \d t ^2 +a^2 (t)  \lp 1+ \frac{2 G m}{a(t) | \vec x |} \rp \d \vec x ^2.
\end{equation}
This metric can be brought in the more familiar form by the coordinate transformations $\vec r = (1+\mu )^2 a(t) \vec x$, meaning \cite{Kaloper:2010ec}
\begin{equation}
a(t) | \vec x | = \frac{G m}{2} \lp \frac{r}{G m} -1 - \sqrt{\lp \frac{r}{G m} -1\rp ^2 -1}\rp ^{-1}
\end{equation}
to get
\begin{equation}\label{met1}
\d s ^2  = - f \d t^2 - \frac{2 H r}{ \sqrt{1-2Gm /r}} \d r \d t +  \frac{\d r^2 }{ 1-2Gm /r} + r^2 \d \Omega_2,
\end{equation}
with $f= 1-2 G m /r - H^2(t) r^2$. In the limit of $a(t) =$ constant or $m\to 0$, one correctly recovers the Schwarzshild or the FRW metric respectively.
It has been shown in Ref. \cite{Kaloper:2010ec}
that, at least in the case ${\rm lim}_{t\to \infty} H(t) \equiv H_0 >0$, obeyed by our universe, the McVittie metric describes a regular (on and outside the horizon) BH embedded in an FRW spacetime with a constant mass.

\begin{figure}
	\centering
	\includegraphics[width=0.55\linewidth]{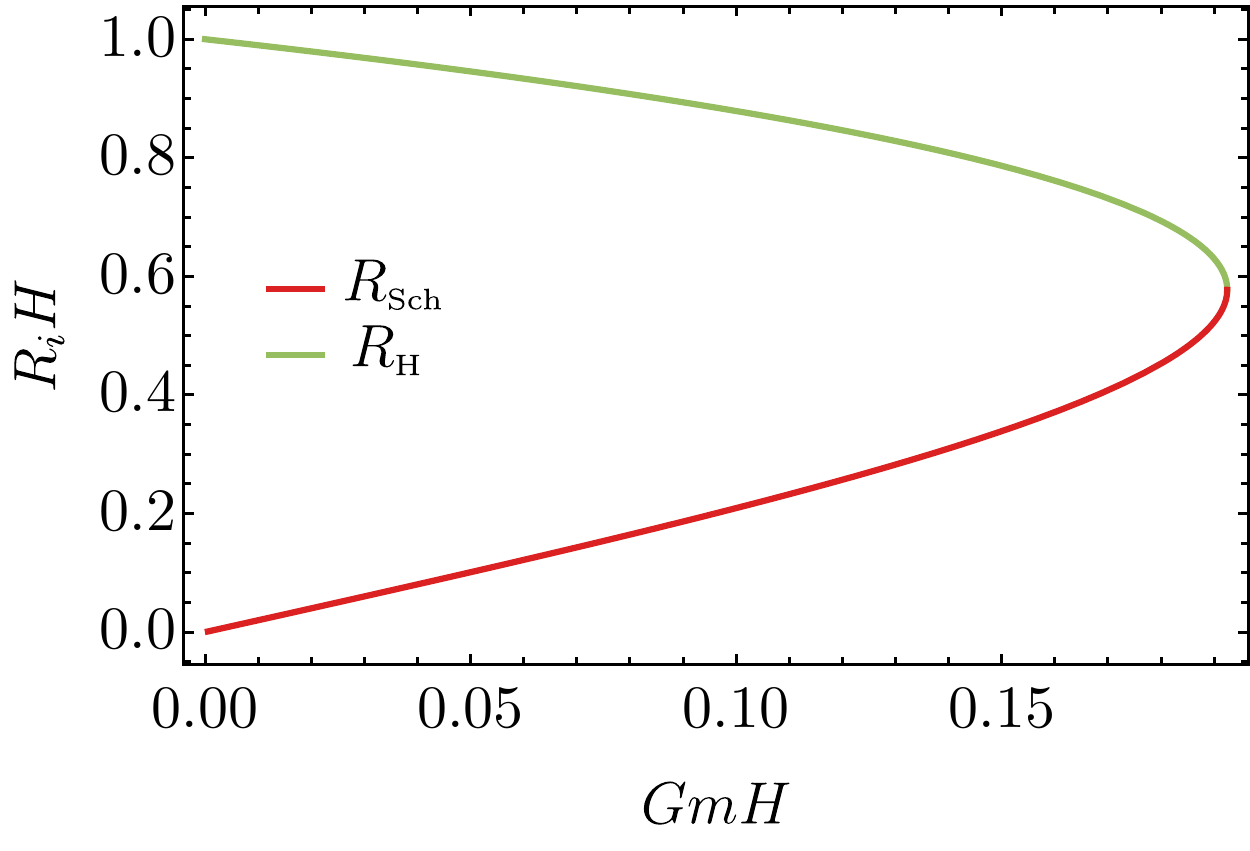}
	\caption{Horizons of the McVittie metric. 
	$R_\text{\tiny H}$ and $R_\text{\tiny Sch}$ are the cosmological horizon and the BH horizon respectively. }
	\label{fig_hor}
\end{figure}

The cross term in Eq.~\eqref{met1} can be eliminated by redefining the time coordinate as
\begin{equation}
\d T = \frac{1}{F} \lp \d t + \beta \d R \rp, 
\end{equation}
where $F(r,t)$ is an integration factor  and 
\begin{equation}
\beta = \frac{H R}{ \lp{1 - 2 G m / R}\rp^{1/2} \lp  1- 2G  m / R - H^2 R^2\rp}, 
\end{equation}
to get
\begin{equation}
\d s ^2 = - \lp 1-2 G m /R - H^2(t) R^2 \rp F^2 \d T^2 +  \frac{\d R^2 }{ 1-2 G m /R - H^2(t) R^2 }+ R^2 \d \Omega_2.
\end{equation}
Therefore, the BH and cosmological horizons are found by solving for the roots of $g^{RR}=0$, giving \cite{Faraoni:2018xwo}
\begin{align}
R_1&= \frac{2}{\sqrt{3} H} \sin \psi, 
\nonumber \\
R_2&= \frac{1}{H} \cos \psi - \frac{2}{\sqrt{3} H} \sin \psi,
\nonumber \\
R_3&= - \frac{1}{H} \cos \psi - \frac{1}{\sqrt{3} H} \sin \psi,
\end{align}
where $\sin (3 \psi) = 3 \sqrt{3} G m H$. Given $m>0$ and $H>0$, $R_3$ is always negative and unphysical, while for $\sin (3 \psi) =1$ one finds the extremal solution where the two horizons coincide. Finally, the BH horizon and cosmological horizon are separated if $\sin(3 \psi)<1$, which is therefore
\begin{equation}
G m H <\frac{1}{3 \sqrt{3}}.
\end{equation}
In such a regime, we can expand the  BH horizon and  the cosmological horizon 
for small values of the mass with respect to the Hubble horizon, meaning the BH horizon is  well within the cosmological horizon, (see also Fig.~\ref{fig_hor} for the exact result)
\begin{align}
R_\text{\tiny Sch}  &= 2 G  m \llp 1+ 4 G^2 m^2 H^2 + {\cal O} ( G^4 m^4 H^4) \rrp  ,
\nonumber \\
R_\text{\tiny H} &=  \frac{1}{H} \llp 1 - G m H - {\cal O} ( G^2 m^2 H^2) \rrp.
\end{align}

In order to gain an intuition on the possible effects of the McVittie metric to the observables usually computed in the Newtonian approximation, we just need to compute the value of the combination $ m H$ entering in the leading order correction to the metric.
As an example, let us consider a PBH with a late time universe mass $m \equiv M_\PBH =  M_\odot$. For that, we can assume it is formed at redshift around 
\begin{equation}
z_\form = 2.2\cdot  10^{12} \lp \frac{M_\PBH}{M_\odot} \rp^{-1/2}.
\end{equation}
Using the fact that, in a radiation-dominated universe, the Hubble rate goes like $H \sim t^{-1} \sim a^{-2}$, we see that rapidly after formation, the McVittie corrections to the ``Newtonian" quantities scales like
\begin{equation}
	G m H \sim \lp \frac{z}{z_\form} \rp^2,
\end{equation}
meaning that the corrections become negligible in the sub-horizon regime within few e-folds after BH formation.
Therefore, one can safely conclude that the GR corrections to the properties of a BH in an expanding universe becomes small
when a hierarchy between the cosmological horizon and the BH horizon is present, i.e. shortly after the BH formation.

We conclude that the LIGO/Virgo bounds are not relaxed by including the effect of the expansion of the universe on the BH metric.


\end{document}